\documentclass[12pt]{article}

\usepackage{sbc-template}
\usepackage{graphicx}
\usepackage{url}
\usepackage[utf8]{inputenc}
\usepackage[english]{babel}
\usepackage{fancyhdr}

\title{A Sustainable Remote Access Architecture for Digital Inclusion through the Reuse of Discredited TV-BOX Devices}

\author{Italo Thiago Felix dos Santos\inst{1}, Carlos Eduardo Correa Queiroz\inst{1}, \\ Adevan Neves Santos\inst{2}, Edgard Luciano Oliveira da Silva\inst{1}}

\address{ Escola Superior de Tecnologia (EST) -- Universidade do Estado do Amazonas (EST/UEA)\inst{1}
\nextinstitute
  Institute of Computing -- Federal University of Amazonas (ICOMP/UFAM)\inst{2}\\
  \email{\{itfds, cecq.eng23, elsilva\}@uea.edu.br, adevan.santos@icomp.ufam.edu.br}
}

\begin{document} 

\maketitle

% Nota de rodapé na primeira página apenas
\thanks{Original title: \textit{Uma Arquitetura de Acesso Remoto Sustentável para Inclusão Digital a Partir do Reaproveitamento de Aparelhos TV-BOX Descaracterizados}. Published in: Proceedings of the XVIII Brazilian Symposium on Ubiquitous and Pervasive Computing 2026, 46th Congress of the Brazilian Computer Society (CSBC 2026), Gramado/RS. Original publication link: \url{https://sol.sbc.org.br/index.php/sbcup/article/view/43317}}

\begin{abstract}
    Digital inclusion and the volume of electronic waste (e-waste) are major challenges for society, with direct impacts on the educational context. Educational institutions, especially those with limited resources, face difficulties in expanding and maintaining computer laboratories. This paper presents the implementation of a sustainable, low-cost Desktop Virtualization Infrastructure (VDI) developed through the reuse of discarded electronic equipment. The proposed solution employs a refurbished Linux server and repurposes confiscated TV Box devices (originally intended for disposal) into functional ARM-based thin clients by installing a compatible Linux distribution. This approach is aligned with the principles of the circular economy by promoting hardware reuse and seeks to reduce the costs associated with deploying computing environments for education. The paper describes the system architecture, the device repurposing process, and the observed results, indicating that this approach represents a viable alternative to support digital education initiatives with a focus on sustainability.
\end{abstract}
     
\begin{resumo} 
    A inclusão digital e o volume de lixo eletrônico (e-lixo) são dois grandes desafios para a sociedade, com impactos também no contexto educacional. Instituições de ensino, especialmente aquelas com recursos limitados, enfrentam dificuldades para ampliar e manter laboratórios de informática. Este artigo apresenta a implementação de uma Infraestrutura de Virtualização de Desktops (VDI) sustentável e de baixo custo, desenvolvida a partir do reaproveitamento de equipamentos eletrônicos descartados. A solução utiliza um servidor Linux recondicionado e converte aparelhos de TV Box apreendidos, originalmente destinados ao descarte, em Thin Clients funcionais baseados em arquitetura ARM, por meio da instalação de uma distribuição Linux compatível. A proposta está alinhada aos princípios da economia circular, ao promover a reutilização de hardware, e busca reduzir custos de implantação de ambientes para ensino de computação. O trabalho descreve a arquitetura, o processo de reaproveitamento dos dispositivos e os resultados observados, indicando que essa abordagem é uma alternativa viável para apoiar iniciativas de educação digital com foco em sustentabilidade.
\end{resumo}

% --- Configuração do Rodapé Global usando FancyHdr ---
\pagestyle{fancy}
\fancyhf{} % Limpa todos os campos (cabeçalho e rodapé)
% Define o rodapé com fonte menor (\footnotesize) e ajusta o espaço
\fancyfoot[C]{{\footnotesize Original Title: Uma Arquitetura de Acesso Remoto Sustentável para Inclusão Digital a Partir do Reaproveitamento de Aparelhos TV-BOX Descaracterizados. Published in: XVIII SBCUP 2026 / CSBC 2026. Available at: \url{https://sol.sbc.org.br/index.php/sbcup/article/view/43317}}}
\renewcommand{\headrulewidth}{0pt} % Remove a linha do cabeçalho se não quiser
\renewcommand{\footrulewidth}{0pt} % Remove a linha do rodapé se não quiser

\section{Introduction}

The Brazilian Federal Revenue Service (RFB) seized approximately 920 million Brazilian reais worth of electronic materials in 2025 \cite{receita-cidada}. Although part of these goods returns to society through online auctions, various items, such as TV Box devices associated with piracy, electronic cigarettes, and beverages, cannot be commercially traded or redistributed directly to the public due to health, safety risks, or intellectual property infringements. The Federal Revenue Service also develops initiatives aimed at disposing of seized merchandise through donations to Civil Society Organizations (CSOs). However, not all data related to seizures and discarded volumes are disclosed publicly in detail. Before the consolidation of programs like Receita Cidadã, much of this material was simply destroyed or discarded, generating significant environmental impact, especially in the case of electronic waste, plastics, batteries, and potentially toxic components. To mitigate this problem, the RFB adopted disqualification policies, a process consisting of the uselessness of brands, logos, and product labels to make them generic. This procedure allows the legal reuse of equipment, enabling their ressignification and subsequent donation to charitable entities, social projects, and educational institutions. This approach gained even more strength during the COVID-19 pandemic \cite{receita-doacao-alcool2020}, when the need to reuse materials became critical. During this period, seized alcohol was converted into alcohol gel for hospital use, and counterfeit or unsuitable masks underwent disqualification and industrial reuse processes, including their transformation into composting inputs and fertilizers. The strategy highlighted the potential to transform products previously destined for destruction into useful resources for society, simultaneously reducing logistical costs and environmental impacts. A recent example of volume occurred in May 2026, with the seizure of 16 tons of electronics at the Port of Rio de Janeiro \cite{receita-noticia-2026}, reinforcing the importance of converting an environmental liability into tools for digital inclusion, research, and technological training.

The democratization of access to technology in educational settings often collides with three main obstacles: the high cost of acquiring hardware, the environmental problem generated by electronic waste disposal, and the recurrent difficulty in institutional mobilization to seek alternatives that overcome dependence on large budgets. The traditional model of computer laboratories, based on individual desktops, requires continuous investment and feeds a cycle of obsolescence and disposal. It is observed that electronic waste disposal centers and technology park renewal processes frequently receive computers and devices that still maintain full functionality but end up underutilized or prematurely discarded. In response to these challenges, this work demonstrates how technical innovation, allied with institutional will, can convert scarce resources into robust educational infrastructures through the circular economy.

The project is based on the implementation of a Virtual Desktop Infrastructure (VDI) in the Digital Systems Laboratory (LSDA/UEA). The initiative takes advantage of the fact that the Federal Revenue Service establishes strategic partnerships with various educational and research institutions for the disqualification and ressignification of smuggled or seized products, which would originally be destined for destruction. The main innovation lies in the transformation of illegal TV Box devices into functional workstations, ressignifying their purpose for education \cite{uea-reuso-2025}. The solution uses a refurbished server that centralizes processing and delivers complete work environments to lightweight terminals, known as Thin Clients.

Recent studies, such as \cite{sabino2017}, have empirically validated the superiority of the Thin Client model compared to the use of TV Boxes as autonomous computers, demonstrating that the VDI approach is the most efficient for extracting performance from these devices in complex tasks. This article, therefore, documents the technical feasibility of an architecture based on a Linux server and communication via the XRDP protocol (X11 Remote Desktop Protocol) \cite{xrdp-server}. The work highlights the potential of this initiative as a replicable model for creating low-cost, high socio-environmental impact laboratories, encouraging other institutions to overcome management barriers and adopt similar measures for sustainable digital inclusion.

% --- CHANGED: Related Works section restructured and enriched ---
\section{Related Work}

The implementation of VDI with free software and low-cost hardware is a consolidated research area. The convergence of this approach with sustainability practices and the circular economy, however, is an emerging field of high social relevance.

\subsection{VDI, Thin Clients, and Performance Analysis}
Free software-based VDI solutions, such as the RDP protocol implemented by XRDP \cite{xrdp-server, microsoft-rdp}, offer robust alternatives to commercial platforms, being frequently applied in university laboratories to optimize resources. The use of Single-Board Computers (SBCs), such as the Raspberry Pi, as Thin Clients is also well documented, demonstrating the viability of reducing workstation costs \cite{morais2019}. When analyzing the capacity of physical servers in VDI environments, Sabino and Carvalho \cite{sabino2017} identified that the evaluated hardware, composed of a server with 8 processing cores and 12 GB of RAM memory, supported between five and six simultaneous users with satisfactory performance. From nine users onwards, the authors observed perceptible degradation of the graphical interface, associated with a total CPU utilization of 94.08\% and only 5.92\% idle time.

Although the authors point out the CPU as the main bottleneck of the system, such results are directly related to the limited computational capacity of the server used and the analysis based on aggregated CPU usage metrics. In scenarios with higher parallelism and a large number of logical cores, such as those evaluated in the present work, the correct interpretation of these metrics becomes fundamental to avoid imprecise conclusions about processing saturation. The study also evidenced the asymmetric nature of network traffic in VDI infrastructures, with average download rates of the order of 1MB/s and upload rates close to 30MB/s, resulting from the continuous transmission of screen updates to terminals. This behavior reinforces the importance of network link quality, especially in educational environments that use wireless connections, an aspect considered in the architecture proposed in this work.

\begin{table}[ht]
\centering
\caption{Technical comparison between VDI architectures and autonomous use.}
\label{tab:comparativo-tecnico}
\small % Slightly reduces font size to fit on the page
\begin{tabular}{|l|c|c|c|c|}
\hline
\textbf{Criterion} & \textbf{Desktop} & \textbf{R. Pi} & \textbf{TV Box (Aut.)} & \textbf{Proposed VDI} \\ \hline
Cost & High & Medium & None/Low & Minimum \\ \hline
Performance & Exc. & Good & M. Limited & High (via Server) \\ \hline
Latency & N/A & Low & N/A & Acceptable (13-25ms) \\ \hline
Scalability & Difficult & Medium & None & High \\ \hline
Sustainability & Low & High & Very High & Very High \\ \hline
\end{tabular}
\end{table}

In this context, the work of \cite{sobrinho2024exploring} is particularly relevant, as it performed a direct comparative performance analysis between a TV Box operating as a personal computer (with Armbian OS) and another as a Thin Client. The results showed that the Thin Client version was drastically superior in all tasks, being up to nine times faster in applications such as video conferencing, a task in which the autonomous version failed due to limitations in graphic processing and decoding. This quantitative evidence not only reinforces the architectural choice of the present project but also justifies the need for a VDI-based infrastructure to extract pedagogical utility from hardware with limited resources. Table \ref{tab:comparativo-tecnico} synthesizes this superiority, confronting the proposed solution with other market architectures. It is observed that while traditional solutions like Desktops offer high performance at high costs and high environmental impact, and devices like the Raspberry Pi have intermediate costs, the TV Box in VDI mode proposal achieves the ideal balance, offering acceptable performance for the school environment with minimum cost and positive environmental impact resulting from the circular economy.

\subsection{Sustainability and National Reuse Initiatives}
The reuse of electronic equipment for educational purposes is a growing initiative in Brazil, aligned with discussions on Green IT policies \cite{camara2022}. A notable example is the project conducted by the Itaipu Technological Park in partnership with the Federal Revenue Service and Unioeste, which transforms seized TV Boxes into mini-PCs for educational and IoT applications \cite{agencia-amazonas-minicomputadores}.

An important aspect of the proposal's sustainability is related to the potential volume of electronic waste avoided by reusing these devices. TV Box appliances have relatively reduced weight; however, the gross weight of the set, comprising the device, power supply, remote control, and HDMI cable, ranges between 300g and 500g. In 2021, the Federal Revenue Service destroyed approximately 97 thousand pirate TV Box devices \cite{agencia-brasil-tvbox2021}. Considering only the conservative average gross weight of 400 g per unit, it is estimated that this operation generated approximately 38.8 tons of electronic waste and associated materials. Even in a more conservative scenario, considering only the average net weight of the devices (approximately 250 g), the discarded volume would still correspond to about 24.25 tons of electronic waste. These values highlight the environmental impact associated with the large-scale destruction of these equipment and reinforce the relevance of technological ressignification initiatives like the one proposed in this work, which seeks to transform part of this environmental liability into low-cost computational infrastructure for digital inclusion and education.

The State University of Amazonas (UEA), also in partnership with the Federal Revenue Service, develops a project that destines ressignified TV Boxes to schools in Baniwa indigenous communities in the Içana river basin \cite{uea-reuso-2025}. Another recent initiative, documented by \cite{daluz2025repurposing}, investigated the reuse of these devices for edge computing applications in smart cities, highlighting their innovative potential and low carbon footprint. These initiatives, including the one presented in this article, demonstrate that the TV Box ressignification model is consolidating as a viable solution to transform an environmental problem into a solution for digital inclusion and education.

\section{System Architecture and Implementation}

The infrastructure was designed in a client-server model \cite{tanenbaum-redes}, where all heavy processing is centralized. This approach is consistent with the literature, which points out the Thin Client model as the most efficient way to extract performance from hardware with limited resources \cite{sobrinho2024exploring}. The core of the infrastructure is a Lenovo ThinkServer TD350 \cite{lenovo-td350}. Reinforcing the commitment to sustainability, the equipment was acquired from an electronic waste recycling and refurbishment center. Equipped with two Intel Xeon E5-2630 v4 processors \cite{intel-xeon-2016} and 160 GB of RAM, the server has only one basic onboard GPU, ASPEED Graphics Family, for console management; it runs Ubuntu 22.04.5 LTS \cite{ubuntu-2204} in "headless" mode. The server has a static IP and connects to the network via Ethernet cable to guarantee maximum performance in delivering remote sessions. 

\begin{figure}[ht]
  \centering
  \includegraphics[width=0.8\textwidth]{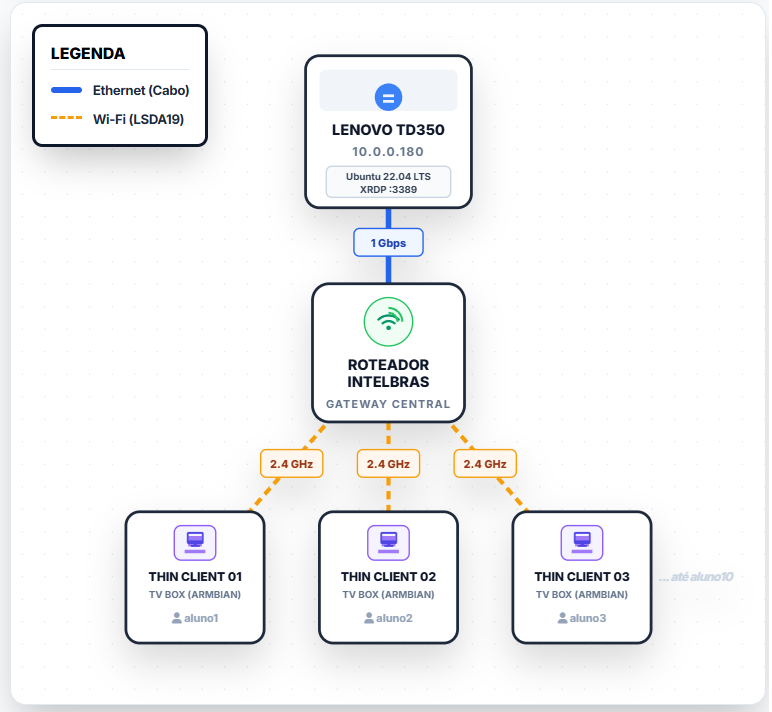}
  \caption{VDI Infrastructure Topology}
  \label{fig:topologia}
\end{figure}

The workstations were created from TV Box devices (model MXQ Pro 4K), originating from Federal Revenue Service seizures. The original Android system was replaced by Armbian Linux \cite{armbian-linux}, a lightweight distribution optimized for the ARM architecture of the devices \cite{armbian-jock}. This modification is the crucial step that transforms a consumer appliance into an educational tool.

\begin{figure}[h]
  \centering
  \includegraphics[width=0.8\textwidth]{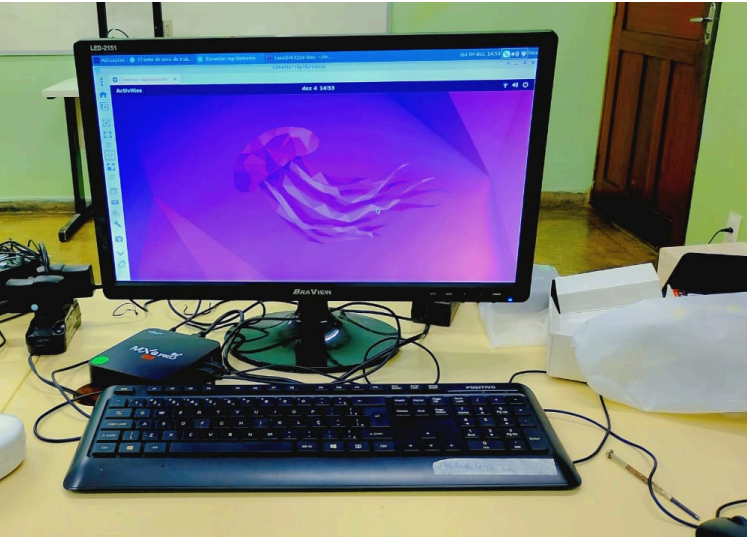}
  \caption{Sustainable workstation with reused TV Box.}
  \label{fig:remmina}
\end{figure}

The Remmina remote desktop client \cite{remmina-client} is used to establish the connection via XRDP with the server, acting as the main interface between the user and the virtualized environment. In this model, the TV Box plays exclusively the role of an access terminal, being responsible only for decoding the video stream transmitted by the server and sending input events from the keyboard and mouse. Thus, the device operates with minimal computational and energy resource consumption, since all heavy processing is centralized on the server. This approach is directly aligned with the principles of Green Information Technology (\textit{Green IT}) \cite{directo2016}, by prolonging the useful life of simple equipment, reducing individual station energy consumption, and minimizing the generation of electronic waste. Figure~\ref{fig:remmina} illustrates one of the stations in operation during tests carried out in the laboratory environment.

\subsection{Management and User Experience}
To unify the user experience, all remote sessions use the GNOME 42.9 environment \cite{gnome-shell-42, desktop-icons-ng}. The interface configuration is standardized and made "immutable" through global rules in the dconf system \cite{gnome-admin-guide}. This ensures that the environment remains consistent and functional for all students, drastically reducing the need for maintenance and technical support.

\section{System Limitations}

Despite the viability of the solution, it is important to highlight technical limitations that primarily affect the reproduction of high frame rate media (video). Initially, unsatisfactory performance in efficiently reproducing videos was attributed to the lack of hardware graphics acceleration (GPU) in the TV Box terminals, whose CPU would be overloaded when decoding the XRDP stream, in line with observations on SBCs operating autonomously \cite{sobrinho2024exploring}. However, subsequent tests conducted with a non-limited hardware Thin Client (a common notebook with dedicated CPU and GPU) connected to the VDI demonstrated that video fluidity remained unsatisfactory and similar to the performance observed in TV Boxes. This evidence indicates that the main performance bottleneck does not reside in the decoding capability of the \textit{client}, but rather in the absence of a dedicated GPU on the server to perform real-time rendering and encoding of this stream.

\begin{figure}[h]
  \centering
  \includegraphics[width=0.8\textwidth]{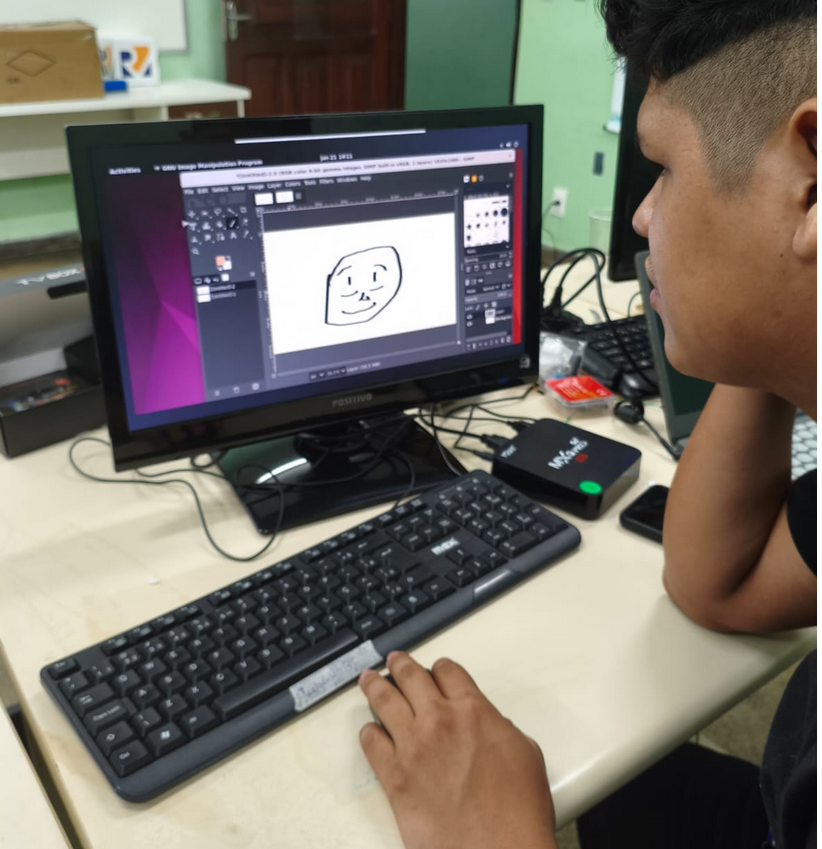}
  \caption{Sustainable workstation with reused TV Box.}
  \label{fig:Gimp}
\end{figure}

Although the Lenovo ThinkServer TD350 server has an onboard GPU, this is intended exclusively for management and console display tasks, not offering effective support for 3D graphics acceleration or hardware video encoding. Thus, graphical rendering of sessions and encoding of the XRDP video stream are performed entirely via software on the server CPU. This characteristic explains the unsatisfactory performance in reproducing media with high frame rates, observed both in TV Box terminals and in clients with dedicated graphics hardware. System monitoring logs indicate high global CPU idle time, combined with significant usage by graphical processes such as \texttt{Xorg} and \texttt{gnome-shell}, a pattern typical of software rendering, confirming that the bottleneck does not reside in the client's capability, but in the lack of adequate graphics acceleration on the server.

\begin{table}[ht]
\centering
\caption{CPU usage distribution for an active user in the VDI infrastructure}
\label{tab:cpu-uso-usuario}
\begin{tabular}{l c}
\hline
\textbf{Metric} & \textbf{Value} \\
\hline
Physical Processors & 2 \\
Total Physical Cores & 20 \\
Total Logical Cores (threads) & 40 \\
Theoretical Max CPU Usage & 4000\% \\
\hline
Typical CPU Usage per Active User (\texttt{top}) & $\sim$110\% \\
Equivalent in Logical Cores & 1.10 \\
Relative Usage of Total Capacity & 2.75\% \\
\hline
Average System Idle Time & 93.4\% \\
Average Total System Usage & 6.6\% \\
\hline
\end{tabular}
\end{table}

For the usage scenario adopted in the laboratory, this limitation is mitigated: typical tasks in the educational environment, such as web browsing, document editing (including the use of GIMP), and simulations (Quartus), function responsively, without perceptible impact on the user experience, even with a high number of simultaneous users, as indicated by the low relative CPU usage rates of the system (Table \ref{tab:cpu-uso-usuario}). Monitors operate with a native resolution of 1280$\times$720 (720p), which also reduces the graphical workload. The ideal solution would be the installation of a virtualization acceleration GPU (\textit{vGPU}) on the server to offload the rendering task, but such cards present high costs and are not easily found. An attempt to use a consumer-grade card (GT 730 4GB) was unfeasible due to the lack of stable drivers for the latest versions of Ubuntu, which prevented CPU \textit{offloading}. Overcoming this bottleneck is an objective for future work.

\begin{table}[h]
\centering
\caption{Perceived Latency as a function of the number of simultaneous users.}
\label{tab:latencia}
\begin{tabular}{|c|c|}
\hline
\textbf{Connected Users} & \textbf{Average Perceived Latency (ms)} \\ \hline
1 & 13 \\ \hline
5 & 13 \\ \hline
10 & 21 \\ \hline
15 & 25 \\ \hline
\end{tabular}
\end{table}

Regarding connectivity, the Wi-Fi interface managed by the Intelbras gateway \cite{intelbras-rede} limits terminals to 100 Mbps \cite{alencar-redes-2010}. Despite this restriction, no network bottlenecks were observed for the intended amount of up to 30 simultaneous accesses. The available bandwidth per student (approx. 3.3 Mbps) is compatible with XRDP session traffic for academic use focused on productivity and software development \cite{tanenbaum-redes}.

\section{Other Limitations}

Although the use of seized TV Boxes depends on specific institutional partnerships, the proposed architecture can be reproduced with other discarded or reused ARM devices, including SBCs, mini-PCs, and embedded equipment from technological disposal. However, the exact reproducibility of the TV Box-based model presents important practical limitations related to hardware heterogeneity. The devices present high hardware heterogeneity, including different processors, board revisions, and memory configurations, which directly impacts compatibility with modern Linux distributions and often requires specific kernels or specialized community support \cite{armbian-boards2026}. In this context, Armbian stands out as a Debian/Ubuntu-based community framework, responsible for providing optimized images and custom kernels for ARM devices, expanding the compatibility and operational stability of these equipment.

The availability of these equipment varies according to the seizure batches by the Federal Revenue Service in each region of the country. Frequently, each regional unit receives unique sets of devices, with distinct characteristics among themselves, making complete infrastructure standardization difficult. Despite this, partner communities and institutions share technical information on compatibility and reuse of these equipment. One of the main repositories used for cataloging and documenting these devices is the \cite{educabox-github2024} project, which gathers information on compatible models, system images, and recovery procedures. Generally, most seized ARM devices can be reused in some computational capacity. However, devices with severe performance, memory, or stability limitations may be destined for less demanding applications, such as automation, embedded controllers, 3D printing, or automated mechanical systems, a strategy also explored by other institutions seeking to extend the useful life of functional hardware, also explored by other institutions linked to \cite{receita-cidada}.

Another relevant factor refers to environmental conditions. In regions with adverse climates, characterized by high temperatures and high humidity, such as the Amazonian region, the impact on the durability of these devices is not yet fully determined. Under these conditions, premature failures related to electronic degradation, oxidation, or thermal instability may occur. Additionally, as reported by the Armbian community \cite{armbian-jock}, some devices may undergo \textit{bricking} or \textit{hard lock} processes, situations in which the equipment becomes permanently unusable, returning to the condition of electronic waste.

Although Table \ref{tab:comparativo-tecnico} presents a comparative analysis of the proposed architecture, it was not possible to perform a precise quantitative financial estimate of the infrastructure. This occurs because a significant part of the computational resources employed consists of refurbished and reused hardware from technological disposal, acquired in periods prior to recent global fluctuations in the semiconductor and electronic component markets. Consequently, absolute cost values might not accurately reflect the current economic scenario. Nevertheless, the continuous rise in hardware prices reinforces the relevance of approaches based on circular economy and technological reuse as sustainable and economically accessible strategies for promoting digital inclusion.

\section{Conclusion}

% --- CHANGED: Stronger conclusion, positioning the work as a confirmation and extension of other studies ---
The results obtained indicate that it is feasible to build low-cost digital inclusion laboratories from the reuse of equipment from electronic disposal. The conversion of TV Boxes into Thin Clients, associated with the use of a refurbished server, demonstrated that a virtualization-based infrastructure can satisfactorily meet the basic demands of educational environments, even in contexts of budgetary restriction. The solution presented stable operation during the testing period and proved suitable for daily academic activities, such as web browsing, document editing, and the use of educational tools, evidencing the potential of the approach to expand access to computational resources. In this context, the work reinforces that the adoption of more sustainable practices in information technology can contribute both to cost reduction and to the decrease in environmental impact, without compromising the minimum usage requirements in teaching environments.

\section{Future Work}

As future work, it is recommended to implement mechanisms for automation of user provisioning and management, in order to reduce administrative effort and increase the reproducibility of the infrastructure. Additionally, it is proposed to investigate optimizations in the network layer to guarantee the quality of user experience (\textit{Quality of Experience} - QoE), especially in scenarios with unstable (\textit{Wi-Fi}) connectivity or high density of simultaneous accesses.

A priority research front, originated from the identified limitations, consists of evaluating the use of dedicated commercial architecture (domestic segment) video cards installed on the server. The objective is to analyze the potential of hardware acceleration for real-time graphical rendering and video encoding in the remote access protocol, overcoming the current software processing bottleneck. This study should identify practical limitations related to driver compatibility in open-source systems, energy consumption, and the impact on the solution's cost-benefit ratio.

Furthermore, the inclusion and evaluation of tools aimed at microelectronics teaching, such as the UNICCASS-ICDESIGN-TOOLS integrated circuit design toolset, is planned. We intend to verify the integration of these tools with the proposed infrastructure and their impact on supporting practical electronic simulation activities. Finally, conducting tests with more demanding engineering applications will allow delimiting the performance limits of the architecture with greater precision and expanding its applicability in different scenarios of higher technology education.

\section*{Acknowledgments}

The authors express their gratitude to the Federal Revenue Service for the fundamental cooperation and for making the TV Box equipment available, which forms the material basis of this research. They extend their thanks to the Vice-Proctorate for Extension (PROEX) of the State University of Amazonas (UEA) % add ufam too?
for fostering the integration between research and extension, through the PADEX program. Finally, they thank the State University of Amazonas for all the institutional support that made this work possible.

\bibliographystyle{sbc}
\bibliography{sbc-template}

@book{tanenbaum-redes,
  author    = {Tanenbaum, Andrew S. and Wetherall, David J.},
  title     = {Redes de Computadores},
  edition   = {5},
  publisher = {Pearson Prentice Hall},
  address   = {São Paulo},
  year      = {2011},
  isbn      = {978-85-7605-924-0}
}

@book{alencar-redes-2010,
  author    = {Alencar, M. A. S.},
  title     = {Fundamentos de Redes de Computadores},
  publisher = {Editora da Universidade Federal do Amazonas (EDUA)},
  address   = {Manaus},
  year      = {2010},
  note      = {Disponível em: \url{https://redeetec.mec.gov.br/}. Acesso em: 13 jan. 2026}
}

@article{daluz2025repurposing,
  author       = {da Luz, Gustavo P. C. P. and Sato, Gabriel Massuyoshi and Gonzalez, Luis Fernando Gomez and Borin, Juliana Freitag},
  title        = {Repurposing of TV boxes for a circular economy in smart cities applications},
  journal      = {Scientific Reports},
  volume       = {15},
  number       = {1},
  pages        = {22638},
  year         = {2025},
  publisher    = {Nature Publishing Group},
  doi          = {10.1038/s41598-025-97379-4}
}

@article{directo2016,
  author    = {Directo, David H. R.},
  title     = {Green Office Computer Workstations Using Thin Client Systems: Energy and Financial Efficiency Study},
  journal   = {International Journal of Smart Grid and Clean Energy},
  volume    = {5},
  number    = {3},
  pages     = {189--196},
  year      = {2016},
  issn      = {2315-4462}
}

@article{sabino2017,
  author  = {Sabino, L. S. and Gontijo, T. S. and Ribeiro, J. M. S.},
  title   = {Análise da Implantação de Virtual Desktop Infrastructures (VDI) em Laboratórios de Informática de uma Universidade},
  journal = {Sistemas e Gestão},
  volume  = {12},
  number  = {4},
  pages   = {436--447},
  year    = {2017},
  doi     = {10.20985/1980-5160.2017.v12n4.1205}
}

@mastersthesis{morais2019,
  author  = {Morais, F. L.},
  title   = {Thin Client Raspberry Pi: Uma Alternativa para Laboratórios de Informática},
  school  = {Faculdade de Tecnologia de Indaiatuba (FATEC)},
  type    = {Trabalho de Conclusão de Curso},
  address = {Indaiatuba, SP},
  year    = {2019}
}

@techreport{sobrinho2024exploring,
  author      = {Sobrinho, M. P. and Sousa, J. V. M. and Carvalho, F. M. P. M. A. and Silva, C. A. N. and Santos, L. R.},
  title       = {Explorando TV Boxes em Ambientes Educacionais: Vantagens e Limitações operando como Thin Client},
  institution = {Universidade Estadual do Piauí (UESPI)},
  year        = {2024},
  address     = {Piripir, PI},
  type        = {Relatório de Pesquisa}
}

@manual{intel-xeon-2016,
  author       = {{Intel Corporation}},
  title        = {Intel® Xeon® Processor E5-2630 v4 Product Specifications},
  year         = {2016},
  organization = {Intel Corporation},
  address      = {Santa Clara, CA},
  url          = {https://ark.intel.com}
}

@misc{lenovo-td350,
  author       = {{Lenovo}},
  title        = {ThinkServer TD350: User Guide and Hardware Maintenance Manual},
  year         = {2025},
  howpublished = {Lenovo Technical Support},
  url          = {https://www.lenovo.com},
  note         = {Acesso em: 05 dez. 2025}
}

@misc{intelbras-rede,
  author       = {{Intelbras}},
  title        = {Equipamentos de rede Intelbras: Manuais de Gateways e Roteadores},
  year         = {2025},
  howpublished = {Portal de Suporte Intelbras},
  url          = {https://www.intelbras.com},
  note         = {Acesso em: 05 dez. 2025}
}

@misc{ubuntu-2204,
  author       = {{Ubuntu Team}},
  title        = {Ubuntu 22.04.5 LTS (Jammy Jellyfish)},
  year         = {2025},
  howpublished = {Canonical Ltd.},
  url          = {https://ubuntu.com},
  note         = {Acesso em: 05 dez. 2025}
}

@misc{gnome-shell-42,
  author       = {{GNOME Foundation}},
  title        = {GNOME Shell: version 42.9 Documentation},
  year         = {2025},
  url          = {https://www.gnome.org},
  note         = {Acesso em: 05 dez. 2025}
}

@misc{desktop-icons-ng,
  author       = {{Desktop Icons NG}},
  title        = {Desktop Icons NG (DING): GNOME Shell Extension},
  year         = {2025},
  howpublished = {GitLab Repository},
  url          = {https://gitlab.com/rastersoft/desktop-icons-ng},
  note         = {Acesso em: 05 dez. 2025}
}

@misc{microsoft-rdp,
  author       = {{Microsoft}},
  title        = {Remote Desktop Protocol (RDP) Specification},
  year         = {2025},
  howpublished = {Microsoft Learn Documentation},
  url          = {https://learn.microsoft.com},
  note         = {Acesso em: 05 dez. 2025}
}

@misc{xrdp-server,
  author       = {{XRDP Project}},
  title        = {XRDP: Open Source Remote Desktop Server},
  year         = {2025},
  url          = {https://www.xrdp.org},
  note         = {Acesso em: 05 dez. 2025}
}

@misc{remmina-client,
  author       = {{Remmina Project}},
  title        = {Remmina: The GTK+ Remote Desktop Client},
  year         = {2025},
  url          = {https://www.remmina.org},
  note         = {Acesso em: 05 dez. 2025}
}

@misc{gnome-admin-guide,
  author       = {{GNOME Project}},
  title        = {GNOME System Administration Guide: Locking down specific settings},
  year         = {2025},
  url          = {https://help.gnome.org/admin/},
  note         = {Acesso em: 05 dez. 2025}
}

@misc{uea-reuso-2025,
  author       = {{UEA}},
  title        = {Projetos da EST/UEA para reuso sustentável de equipamentos apreendidos pela Receita Federal ganham destaque nacional},
  year         = {2025},
  howpublished = {Portal UEA},
  url          = {https://www.uea.edu.br/index.php/2025/03/20/projetos-da-est-uea/},
  note         = {Acesso em: 05 dez. 2025}
}

@misc{agencia-amazonas-minicomputadores,
  author       = {{Agência Amazonas}},
  title        = {Secretaria de Educação recebe 25 mil minicomputadores para ampliar inclusão digital},
  year         = {2025},
  howpublished = {Portal do Governo do Estado do Amazonas},
  url          = {https://www.agenciaamazonas.am.gov.br/},
  note         = {Acesso em: 05 dez. 2025}
}

@misc{armbian-jock,
  author       = {{jock}},
  title        = {RK322X: Support for generic RK3228A, RK3228B and RK3229 TV boxes},
  year         = {2025},
  howpublished = {Armbian Community Documentation},
  url          = {https://www.armbian.com/author/jock/},
  note         = {Acesso em: 13 jan. 2026}
}

@misc{armbian-boards2026,
  author = {{Armbian}},
  title = {Documentação e suporte a placas suportadas pelo Armbian},
  year = {2026},
  howpublished = {\url{https://armbian.com/pt/boards/}},
  note = {Acessado em: 1 de maio de 2026}
}

@misc{camara2022,
  author       = {{Brasil}},
  title        = {Projeto de Lei nº 587, de 15 de março de 2022: Institui a Política Federal TI Verde},
  year         = {2022},
  howpublished = {Câmara dos Deputados},
  url          = {https://www.camara.leg.br/proposicoesWeb/},
  note         = {Acesso em: 01 mai. 2024}
}

@misc{armbian-linux,
  author = {{Armbian Team}},
  title  = {Armbian Linux: The computing framework you can rely on},
  year   = {2025},
  url    = {https://www.armbian.com},
  note   = {Acesso em: 05 dez. 2025}
}

@misc{receita-cidada,
  author = {{Receita Federal}},
  title = {Receita Cidadã: Destinação Sustentável},
  year = {2026},
  howpublished = {\url{https://www.gov.br/receitafederal/pt-br/acesso-a-informacao/acoes-e-programas/cidadania-fiscal/novos-destinos}},
  note = {Acessado em: 15 de maio de 2026}
}

@misc{receita-noticia-2026,
  author = {{Receita Federal}},
  title = {Receita Federal apreende cerca de 16 toneladas de produtos eletrônicos irregulares no Porto do Rio de Janeiro},
  year = {2026},
  howpublished = {\url{https://www.gov.br/receitafederal/pt-br/assuntos/noticias/2026/maio/receita-federal-apreende-cerca-de-16-toneladas-de-produtos//
  -eletronicos-irregulares-no-porto-do-rio-de-janeiro}},
  note = {Acessado em: 13 de maio de 2026}
}

@misc{educabox-github2024,
  author = {{EducaBox202}},
  title = {Repositório Oficial do Projeto EducaBox},
  year = {2024},
  howpublished = {\url{https://github.com/educabox/educabox}},
  note = {Acessado em: 15 de maio de 2026}
}

@misc{receita-doacao-alcool2020,
  author = {{Receita Federal}},
  title = {Receita Federal doa 30 mil litros de bebidas alcoólicas para fabricação de álcool em gel},
  year = {2020},
  howpublished = {\url{https://www.gov.br/receitafederal/pt-br/assuntos/noticias/2020/abril/receita-federal-doa-30-mil-litros-de-bebidas-alcoolicas-para//
  -fabricacao-de-alcool-em-gel}},
  note = {Acessado em: 13 de maio de 2026}
}

@misc{agencia-brasil-tvbox2021,
  author = {{Agência Brasil}},
  title = {Receita Federal destrói 97 mil aparelhos de TV Box piratas},
  year = {2021},
  howpublished = {\url{https://agenciabrasil.ebc.com.br/economia/noticia/2021-05/receita-federal-destroi-97-mil-aparelhos-de-tv-box-piratas}},
  note = {Acessado em: 13 de maio de 2026}
}

\end{document}